\documentclass[12pt]{article}
\usepackage[figuresright]{rotating}
\usepackage{graphics}
\usepackage{graphicx}
\usepackage{epstopdf}
\usepackage{adjustbox}
\usepackage{tabulary}
\usepackage{hyperref}
\usepackage{authblk}
\usepackage{tabularx}
\usepackage[font=small,skip=0pt]{caption}
\usepackage{subfigure}
\usepackage{verbatim}
\usepackage{amsmath, amssymb}
\usepackage{amsthm}
\title{From Public Outrage to the Burst of Public Violence: An Epidemic-Like Model}

\author[1, 2]{Sarwat Nizamani\thanks{saniz@mmmi.sdu.dk}}
\author[1]{Nasrullah Memon\thanks{memon@mmmi.sdu.dk}}
\author[3]{Serge Galam\thanks{serge.galam@cnrs-bellevue.fr}}

\affil[1]{The Maersk McKinney Moller Institute,University of Southern, Denmark}
\affil[2]{University of Sindh, Pakistan}
\affil[3]{Centre National de la Recherche Scientifique (CNRS), France}

\date{(October 2, 2013)}
\begin{document}

\maketitle

\begin{abstract}
This study extends classical models of spreading epidemics to describe the phenomenon of contagious public outrage, which eventually leads to the spread of violence following a disclosure of some unpopular political decisions and/or activity. Accordingly, a mathematical model is proposed to simulate from the start, the internal dynamics by which an external event is turned into internal violence within a population. Five kinds of agents are considered: ``Upset'' (U), ``Violent'' (V), ``Sensitive'' (S), ``Immune'' (I), and ``Relaxed'' (R), leading to a set of ordinary differential equations, which in turn yield the dynamics of spreading of each type of agents among the population. The process is stopped with the deactivation of the associated issue. Conditions coinciding with a twofold spreading of public violence are singled out. The results shed a new light to understand terror activity and provides some hint on how to curb the spreading of violence within population globally sensitive to specific world issues. Recent world violent events are discussed.

\end{abstract}
\textbf{Keywords:}  sociophysics, opinion dynamics, epidemic modeling, public violence, minority spreading


\section{Introduction}
The topic of epidemic spreading has inspired a great deal of research for many years, leading to the development of a series of models \cite{14},\cite{29} and \cite{30} for which the SIR is a cornerstone \cite{14}. One main focus is the determination of the existence of a threshold for the spread in parallel to evaluating the time scale of the epidemic. Indeed, the spreading of an illness obeys similar qualities as the spreading of minority opinions due to the fact that both are based on local interactions among a few agents starting from at least two. The rumor-spreading phenomenon is most emblematic of the analogy \cite{10}, \cite{11} while models of epidemics are continuously using ordinary differential equations \cite{14}, opinion dynamics models use either continuous \cite{31} or discrete variables \cite{32}. 
\\
In this analysis, we extended the application of the mathematical frame of ordinary differential equations with continuous variables to investigate the propagation of hatred within a subclass of a heterogeneous population. In particular, we focused on the conditions by which the spreading of public outrage leads to outbreaks of public violence.
\\
We focused on delayed reactions of outrage, which occur in one part of the world, driven by happenings which took place earlier in a different part of the world. More precisely, we concentrate on political actions that were perceived locally to be harmless and insignificant while perceived by distant populations as unbearable offense \cite{1, 2, 3}.
\\
For instance, we can cite the making of the Anti-Islam movie named ``Innocence of Muslims''\footnote{http://en.wikipedia.org/wiki/Reactions-to-Innocence-of-Muslims}  by an Egyptian-born US resident\footnote{http://en.wikipedia.org/wiki/Nakoula-Basseley-Nakoula}.  When an Arabic-dubbed version was put on air in September 2012, it took few weeks before reactions flared outside the US, leading to the death of dozens of people and hundreds of injuries. Another example took place in India; known as ``Operation Blue Star 1984,'' a military operation on the sacred grounds of one of Sikhs community in India\footnote{http://en.wikipedia.org/wiki/Operation-Blue-Star} caused around 5000 casualties.
\\
Such remote reactions \cite{1, 2, 3} may last for days, weeks, or months, creating fear and hatred among people of different nations. Subsequent violent incidents are considered to be ``hate crimes'' \cite{7} also known as ``hate violence'' \cite{8}.
\\
However, the same kind of ``provocation'' does not automatically lead to a burst of transnational public violence, as shown with the publication of caricatures of the Muslim's Prophet by French journalist Charlie Hebdo\footnote{http://www.france24.com/en/20130102-french-satire-publishes-life-mohammed-cartoon-charlie-hebdo}. 
There are two different states of ``contamination'' in the model. The first state is an agent being ``Upset'' (U) by the inciting event and the second step is an agent turning ``Violent'' (V) to implement revenge. In addition, following the epidemic nomenclature, we introduce Sensitive agents (S), Immune agents (I) and Relaxed agents (R)   leading to a total of five kinds of agents.
\\
The rest of the paper is organized as follows: Section 2 reviews a series of related works, while Section 3 presents our model combining the five kinds of agents with the rules of interactions. The model equations are solved numerically in Section 4 for a series of specific cases, following the variations of proportions of each kind of agents. Section 5 describes the various stages of the ending dynamics. Lastly, Section 6 concludes the paper mentioning possible directions for future work. 

\section{Related Work}
In this paper, we present a mathematical model for hatred issue awareness, which to the best of our knowledge is the first attempt to mathematically model such issues. We discuss related work in connection to proposed model in this section, which are epidemic models or rumor-spreading models. The epidemic model \cite{14} describes a general framework, which is then extended to rumor spreading. According to basic framework, a population of N individuals is divided into three groups. The process of the epidemic begins when an individual gets an infection and turns into an infectious (I) agent. The rest of the agents in the population become susceptible (S) to the infection and are referred as S. The infectious agents may spread the infection among S by contacting them. The agents I who recover and do not get re-infected nor spread the infection are referred as R agents. The basic rumor spreading model is comprised of the three states of the epidemic model, which was first proposed by Daley and Kendal \cite{10}, and is referred to as the DK model. The general framework of the DK model considers N as the total population size, which is divided in three types of agents, namely: the agents S are the individuals who are aware of the rumor and willing to spread it; the agents I are the individuals who are ignorant about the rumor and are susceptible to rumor; while R are the individuals who know about the rumor but are not interested in spreading it and are referred to as stiflers. According to the model, when spreader meets ignorant, the ignorant becomes the spreader. If spreader meets another spreader or the stifler, then the spreader turns into the stifler. 
\\
A variant of basic DK model was later proposed, called the MK \cite{11} model, which says that when spreader meets ignorant, then ignorant turns into spreader; when spreader meets another spreader, one of them turns into stifler; with the interaction of the stifler to the spreader, turns spreader into stifler. In both DK and MK models, ignorants only decrease over time. Both of these models use three agent categories of population S, I, and R and present a mathematical theory of spreading mechanism of epidemic diseases \cite{14}. As in epidemic disease, when an individual gets infected, the rest of the population becomes susceptible to the disease based on contact. This basic model was then used and extended for rumor spreading in numerous studies \cite{10, 11, 12, 15, 16, 17, 18}. 
\\
Zenette \cite{17} studied the dynamics of rumor propagation on small-world networks by considering these three base agent types of rumor spreading.  The study \cite{19} presents a rumor-spreading model in a slightly different perspective, which shows how a minority, believing the rumor to be the truth, changes the majority to accept the rumor in a very short period of time. The study suggests that while accepting the rumor, the bias of the individuals towards the subject of the rumor has an important effect.
\\
Zhao et al. \cite{12} presented their SIHR model that extended the SIR rumor-spreading model by introducing a new state of hibernator, which is comprised of the spreaders who have forgotten the rumor for the time being. The hibernators then move back to spreader state by remembering the rumor. In another study, Zhao et al. \cite{20} have extended the SIHR model by showing the behavior of rumor spreading in populations of inhomogeneous networks. The most recent study \cite{18} explores the rumor spreading with another pattern, i.e., ISRW, which considers the medium as subclass (an agent) to which the rumor can be transmitted from and vice versa. 
\\
Apart from these rumor-spreading and epidemic models, a number of studies \cite{21, 22, 23, 24, 25} describe various social phenomena using aspects of physics. Galam \cite{22} describes global terrorism using percolation theory and the study deduced that in the current situation of global terrorism in which there passive supporters of terrorism, the military actions are insufficient. Galam and Mauger\cite{21} present a model based on percolation theory [26] to describe global terrorism and suggests global terrorism can be reduced by decreasing percolation dimensions. The study inspired us to model such an important global social problem, whose reactions are very destructive.
\\ 
The proposed model extends the basic SIR framework in the context of hatred issue-awareness with additional two types of agents. In total, five types of agents are considered with the following distinguished roles:
\begin{itemize}

\item The agents S are sensitive persons who are un-aware of the issue and belong to the issue-sensitive population.  These agents have a similar role as ignorant /susceptible in rumor spreading / epidemic models.
\item The agents U are upset persons who are aware of the issue and are involved in spreading it by using some communication media (for example social media) or social gatherings, but do not take any violent action against the issue.  These are like spreaders/infectious in rumor spreading / epidemic models. 
\item	The agents I are immune agents, who have their viewpoint for which they do not get upset, like vaccinated agents in SIR model. However, the I agents are contacted by the U agents to get them upset like the failure of vaccination in epidemic model.
\item	The agents V are the violent persons who are involved in violent reactions against the issue. The reactions may be in the form of protests or targeting specific individuals or locations and have the highest probability of inflicting injury or death. 
\item	The agents R are the relaxed, who become relaxed after staying upset for $\frac{1}{\xi}$ unit times and violent for $\frac{1}{\eta}$ unit times from the agent U and the agent V, respectively.  These agents are also the additional agents in the model, who neutralize agents U and V; when everyone in U and V groups turn to agent R, the issue vanishes. At the end of the issue, the agents R and I are alike; however, before the end of the issue, the I agents are contacted by the agents U and some of them turn to U. That is why the two agent types are considered separately.
\end{itemize}
The proposed model is discussed in Section 3, which simulates the dynamics of drastic issues by answering the following research questions: 
\\
Q1. Does the mathematical model have the potential to describe the effect of the inciting incidents?
\\
Q2.What is the effect of interaction between various population categories during the lifetime of violence-causing events?
\\
Q3. What is the final state of the issue at the end?
\\
In order to address the research questions, the following hypotheses are designed and supported by the proposed model:
\\
H1. The results of inciting incidents can be well-described mathematically using mean-field equations.
\\ 
H2. The interaction among various issue sensitive populations causes increase in violent and upset agents in the beginning, then decrease with the passage of time.
\\
H3. The issues vanish at the end. 
\\
In the following section, we discuss the proposed model of hatred issue awareness beginning with the basic SIR framework [14] of three types of the agents, then continue to an extended model comprised of four types of agents and leading to a proposed model of five types of agents.

\section{Proposed Model}
The proposed model encompasses five types of agents, an expansion of the basic epidemic model which leads to hatred contamination, then from hatred contamination to violence, ushering in a five-type agent model.
\subsection{From epidemics to hatred contagion }
As in epidemiology where the epidemic process is initiated when an infectious agent contaminates other agents, hatred can be passed by exchanging words, whether fact or rumor. However, in both cases, only susceptible agents become contaminated. A spreading process is thus initiated which can either lead to a large-scale epidemic or simply fade away. 
\\
Nevertheless, in the case of hatred, besides getting upset, an agent can turn violent, creating destruction outside the pure process of being outraged. Such a state seems to be specific to the case of hatred. At the same time, some agents are immune towards the emotional content of the incriminated issue.
\\
To set our model, we started from an analogy with a basic SIR epidemic model \cite{14} by considering three types of agents: Sensitive (S), similar to susceptible agents, Upset (U), similar to infectious agents, and Relaxed (R), equivalent to recovered agents. The simplest form of the model is shown in Figure 1 where each square represents an agent type, while an edge shows the transition between agent types showing the rate at which transition takes place.

\begin{figure}[h]
\begin{center}
\includegraphics[scale=0.5]{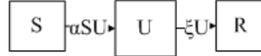}
\caption{SIR-like simple model}
\end{center}
\end{figure}

Similarly, to many diffusive phenomena, only a few agents are upset at the beginning. However, once upset, these agents come into contact with sensitive agents trying to make them upset too, turning on the process of hatred spreading. The transition from S to U depends on contact rate per unit time, resulting in reproduced U agents. It is given as $ \alpha P U \left(\frac{S}{P} \right)= \alpha U S $(where P is the total sensitive population), the number of new U agents per unit time reproduced. Accordingly, agents S turn to agents U at a rate $\alpha$, producing an increase of $\alpha U S $ in the U agents.  At the same time, people do not stay upset forever about a specific issue and eventually relax. We assume it occurs at a constant rate $ \xi$, i.e., on average a U individual is relaxed after $ \frac{1}{\xi}$ time units with relaxing transmission represented as $\xi U $. An associated process is described by the set of ordinary differential equations,

\begin{equation}
\frac{dS}{dt}=-\alpha S U 
\end{equation}
\begin{equation}
\frac{dU}{dt}=\alpha S U - \xi U
\end{equation}
\begin{equation}
\frac{dR}{dt}=\xi U
\end{equation}

On this basis, we add the possibility for an S agent in contact with a U agent not to get upset and instead become immune to the issue, similar to the case of either natural immunity or vaccination in epidemics. The proportion of I agents produced per unit time is given by $ \beta P U  \left( \frac{S}{P}\right)  = \beta U S $ leading to the equations,
\begin{equation}
\frac{dS}{dt}=-\alpha S U-\beta S U
\end{equation}
\begin{equation}
\frac{dI}{dt}=\beta S U
\end{equation}
as shown in Figure 2.
\begin{figure}[h]
\begin{center}
\includegraphics[scale=0.5]{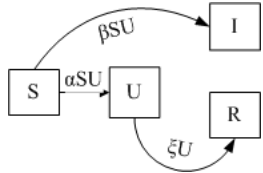}
\caption{Addition of I (immune) agent type}
\end{center}
\end{figure}
\\
However, it is known in epidemiology that vaccination does not always produce 100\% immunization. The same holds true here with the possibility for an I agent to eventually turn upset while having a subsequent contact with the U agents as shown in Figure 3. The rate of shift per unit time is denoted by Eqs. 2 and 5 and are modified respectively as, 
\begin{equation}
\frac{dU}{dt}=\alpha S U - \xi U +\kappa U I
\end{equation}
\begin{equation}
\frac{dI}{dt}=\beta S U - \kappa U I
\end{equation}

\begin{figure}[h]
\begin{center}
\includegraphics[scale=0.5]{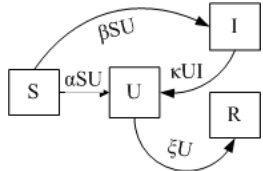}
\caption{ Addition of a new edge I to U in result of some failed immunization}
\end{center}
\end{figure}
\subsection{From hatred to violence}
Now we discuss an important aspect of the hatred issue, in which some of the sensitive agents are extremely sensitive; in addition to becoming upset, they turn violent in order to implement some revenge for what they considered as an outrage. They are denoted as violent agents (V). Two paths to create a V agent are introduced. A direct path results from an encounter between U and S agents at a rate $ \gamma $ while the second one emerges from encounters between two U agents at a rate $ \sigma $. In addition, V agents can also turn S agents into U at a rate $ \mu $. V agents also relax after $\frac{1}{\eta}$ unit times. 
\\
A new differential equation is thus obtained for the V agent dynamics in addition to a modification of Eqs. 3, 4, and 6, which yield,
\begin{equation}
\frac{dS}{dt}=-\alpha S U-\beta S U-\gamma S U-\mu S V
\end{equation}
\begin{equation}
\frac{dU}{dt}=\alpha S U-\xi U+\kappa S I-\sigma U U+\mu S V
\end{equation}
\begin{equation}
\frac{dR}{dt}=\xi U+\eta V
\end{equation}
\begin{equation}
\frac{dV}{dt}=\gamma S U-\eta V+\sigma U U
\end{equation}
In the final set, we have a set of five differential equations to describe the dynamics of the respective proportion of each kind of agent, i.e., sensitive (S), upset (U), immune (I), relaxed (R), and violent (V) as illustrated in Figure 4.
\begin{figure}[h]
\begin{center}
\includegraphics[scale=0.5]{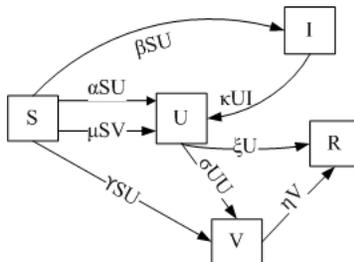}
\caption{The model with five agent types}
\end{center}
\end{figure}
\subsection{Implementing the five-agent model }
To begin implementing the model, we start with some initial conditions ($ S_{0}, U_{0}, I_{0}, V_{0}, R_{0} $) at time t = 0 under the constraint $ S_{0} + U_{0} + I_{0} + V_{0} + R_{0} = 1 $, to study the various possible scenarios of the time evolution of those proportions ($ S_{t}, U_{t}, I_{t}, V_{t}, R_{t} $) at time t. 
Indeed, when a bursting event occurs somewhere in the world, focusing in some specific part of the world where the population is sensitive to such an event, at t = 0 we have only a few sensitive agents who become aware of it. Subsequently, they get either upset with a proportion $ U_{0} $ or immune with proportion $I_{0} $. The rest of the population stays in the sensitive state with $ S_{0} =1 - U_{0} - I_{0} $ with $ V_{0} = R_{0} = 0 $. Those initial values result from an external effect. Then, internal dynamics are turned on, driven by a word-of-mouth phenomenon. 
Hence, the lifetime internal dynamics take on a shape of a system of linear equations given in the matrix form:
\\
\begin{center}
$ \begin{matrix}t_{0}\\t_{1}\\.\\.\\t_{\infty} \end{matrix}
\begin{bmatrix}
S_{0} & U_{0} & 0 & 0 & I_{0}\\
. & . & . & . & . \\
. & . & . & . & . \\
. & . & . & . & . \\
S_{\infty} & 0 & 0 & R_{\infty}& I_{\infty} \\
\end{bmatrix} 
=\begin{bmatrix} 1 \\ 1 \\ 1 \\ 1 \\ 1 \\\end{bmatrix} $ 
\end{center}

The internal dynamics stop at a certain time, $t = \infty $, resulting in the steady state of the system, where hatred and violence are no more active. The internal dynamics consequences in the final assumptions $ S_{\infty} = 1 - R_{\infty} - I_{\infty} $. Thus, $R_{\infty} + I_{\infty}$ yields the final number who ever was aware of the ``bursting'' event.
\section{Numerical Simulations}
In order to numerically analyze the proposed model, we consider the parameter values and their description given in Table 1.

\begin{table}
\centering
\begin{tabular}{|l|l|l|}
\hline
$ \alpha $ & U/S contacts per unit time to yield U agents &  0.5-5 \\ \hline
$\beta $ & U/S contacts per unit time to yield immune agents & 0.5 \\  \hline
$\gamma $& U/S contacts per unit time at transmission rate $\gamma $ from S to V &  0.2 \\  \hline
$\mu $& V/S contacts per unit time at transmission rate $\mu $  from S to U  & 0.1 \\ \hline
$\xi$ & 1/$\xi$ time units an individual remains spreader & 1-30 \\  \hline
$\eta $& 1/$\eta $ time an individual remains violent & 1-20 \\ \hline
$\kappa$ & Contact rate that transmits I to U & 0.5 \\ \hline
$\sigma$ &  U/U contact rate transmitting them to V&  0.5 \\ \hline
\end{tabular}
\caption {Model parameters with description and values} 
\end{table}

\subsection{From epidemic to hatred contagion}
 The numerical simulations of this simple model illustrate the behavior of the model in the simple case when we have only three types of agents and only two transitions. As the agent type S is monotonically decreasing and the agent type U is increasing by receiving agents from the S and then decreasing by transmitting the agents to the agent type R. In the beginning, the density of the U increases, but after reaching a peak value, it begins to decrease. We needed to determine the necessary conditions for U to increase and then to decrease. We also determined the condition for U to outbreak (necessary condition for which U increase). The basic condition for U to increase is that when $ S_{t} > \frac{\xi}{\alpha} $, while $ S_{t} < \frac{\xi}{\alpha} $ causes U to decrease. It is a necessary condition for outbreak that $ S_{0} > \frac{\xi}{\alpha} $, hence U will increase up until some point.  Another condition for outbreak is to determine reproduction rate  \cite{28}  $ \textit{\c{R}} > 1 ( \textit{\c{R}}  $ is a reproduction rate and it should not be confused with agent type R). The reproduction rate determines the number of secondary U agents produced by the initial U agents. It is determined as $\textit{\c{R}} =\frac{\alpha}{\xi} $, if the condition holds, the issue may be considered serious. Figure 5 (a-e) illustrates the basic model of hatred contagion by varying parameter values; hence the condition of peak value can also be verified.

\begin{figure}
 \centering
 \subfigure[Peak value of U at $ t=11.12$ time units  and $S_{t}\approx \frac{\xi}{\alpha}=0.222$]{
    \includegraphics[width =50mm] {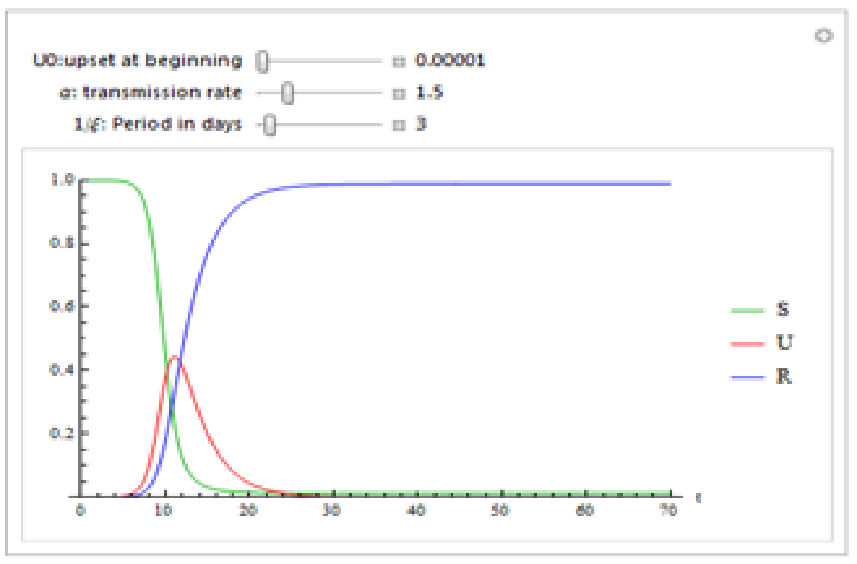}
  }
  \subfigure[ Peak value of U at $ t=5.058 $ time units, and $S_{t}\approx \frac{\xi}{\alpha}=0.05 $]{
      \includegraphics[width =50mm] {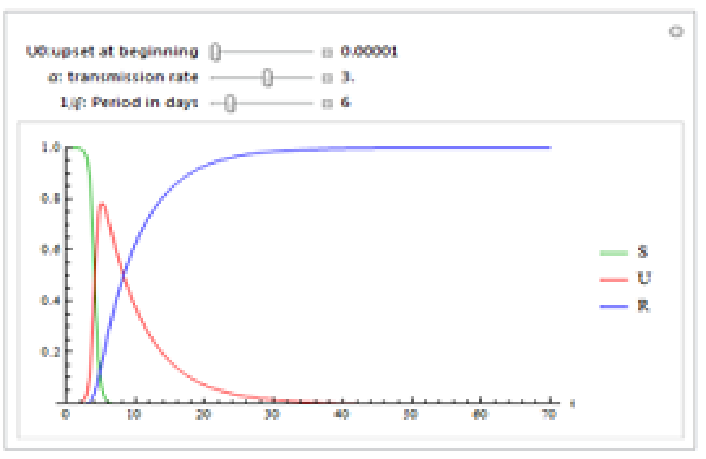}
    }    
   \subfigure[ Peak value of U at $ t=5.273$ time units, and $S_{t}\approx \frac{\xi}{\alpha}=0.111 $]{
        \includegraphics[width =60mm] {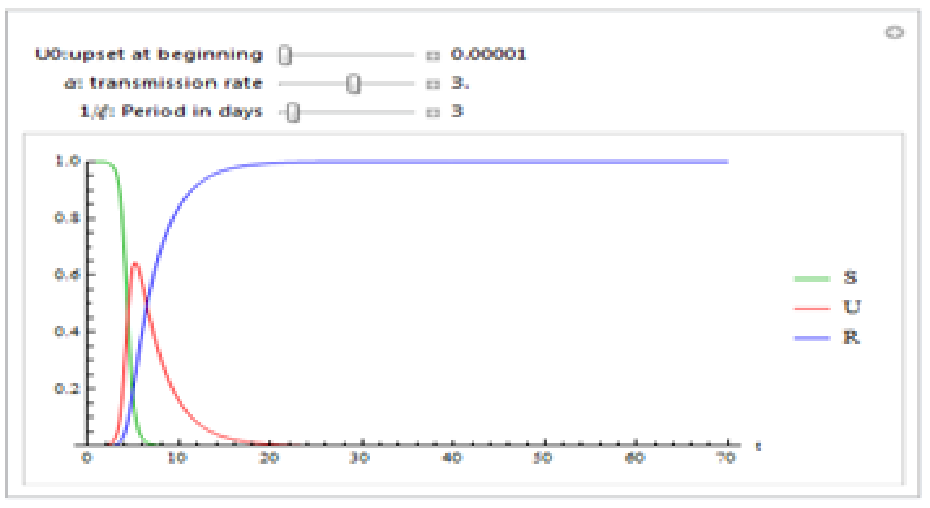}
        }
        
         \subfigure[Peak value of U at $ t=20.56 $ time units, and  $S_{t}\approx \frac{\xi}{\alpha}=0.208$]{
                \includegraphics[width =55mm] {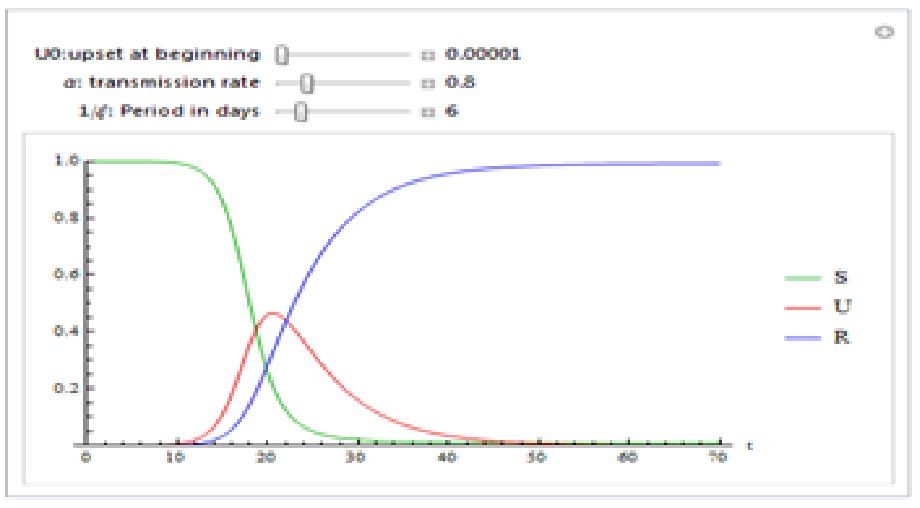}
                }
                  \subfigure[Peak value of U at $ t=22.249 $ time units and  $S_{t}\approx \frac{\xi}{\alpha}=0.3125$]{
                                \includegraphics[width =50mm] {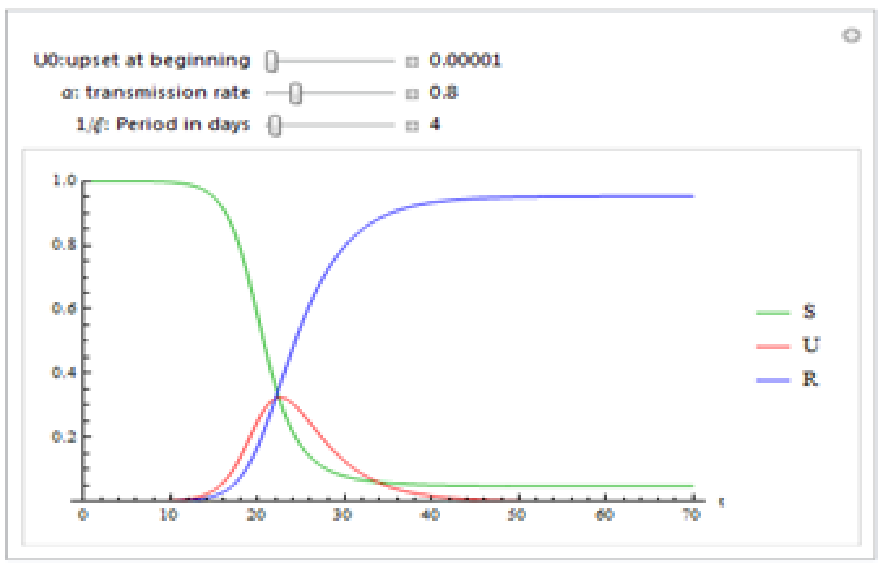}
                                }
 \caption{Basic model with varying range of parameters}
\end{figure}

For numerical simulations we have used varying values of the parameters ($\alpha$ and $\xi$) in Figure 5 (a-e). In each of the Figure peak value of the U agents has been given with the necessary condition for outbreak for the given peak.

\subsection{Hatred contagion with immunization}
An enhancement to the basic model has been integrated by introducing the I (immune) agent type, which is resulted by the individual's viewpoint to not get upset. Figure 6 shows the dynamics of the whole system with newly introduced agent type for varying values of parameters. In this enhanced model, the reproduction rate depends on more parameters. It is estimated as $\textit{\c{R}} = \frac{\alpha+\kappa}{\beta+\xi} $ and if $ \textit{\c{R}} > 1 $ , then outbreak is likely to occur, meaning a significant number of secondary U agents will be produced. As $\alpha$ is the rate at which the agents S turn to agent U and $ \kappa $ is the rate at which the I agents (losing their immunization) turn to the U agents; while at the rate of $ \beta $, the agents S are immunized (do not turn U) and $ \xi $ is the rate at which the agents U turn relaxed. Hence, the reproduction rate $ \textit{\c{R}} >1 $ lets outbreak for the agent U to occur. In Figure 6 (a-d), the dynamics of the agents have been illustrated and also the reproduction rate is given for each of the varying values of $\alpha $ and $ \xi $.

\begin{figure}
 \centering
 \subfigure[\textit{\c{R}} = 2.4]{
    \includegraphics[width =55mm] {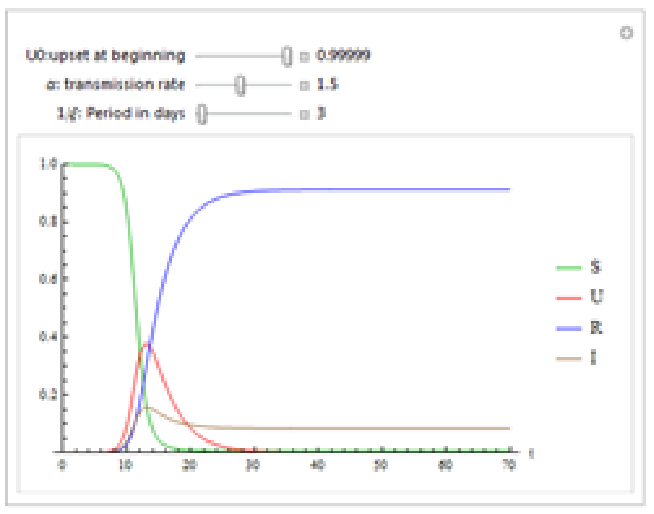}
  }
    \subfigure[\textit{\c{R}} = 3.55385]{
      \includegraphics[width =55mm] {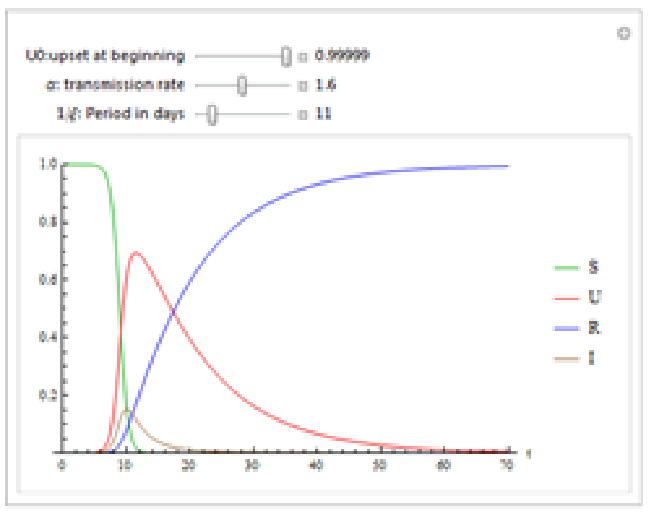}
    }
       \subfigure[\textit{\c{R}} = 2.85]{
        \includegraphics[width =55mm] {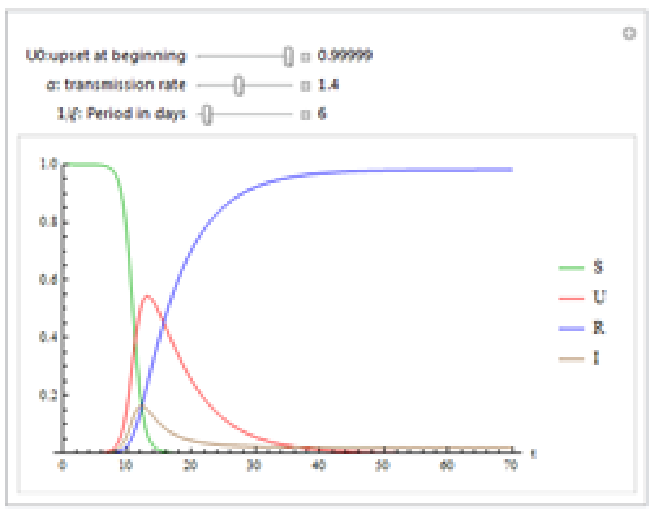}
        }
                 \subfigure[\textit{\c{R}} = 1.32]{
                \includegraphics[width =55mm] {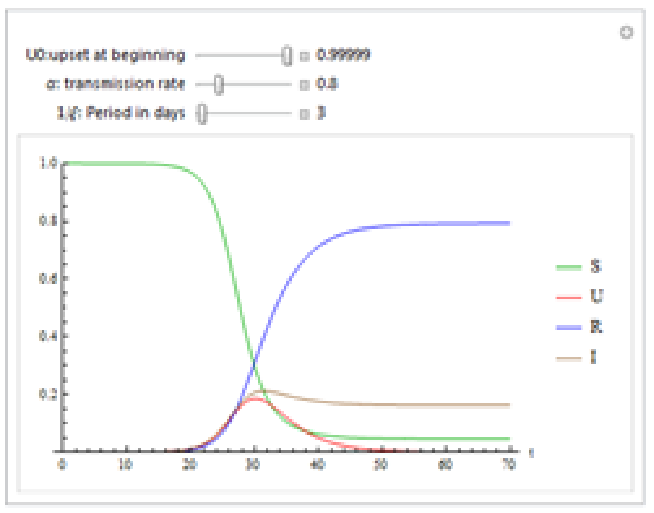}
                }
 \caption{Numerical simulations of extended model with Immunization}
\end{figure}

In Figure 6 (a-d), we presented dynamics of the issue awareness with extended model. In this model we estimated reproduction rate, which determines whether there will be outbreak for the U agents or not. In each illustration this reproduction rate has also been given. It should be observed that higher is the reproduction rate, the maximum peak the U agents attain. 

\subsection{From hatred to violence}
A fully evolved proposed model for hatred issue awareness is comprised of 5 types of agents and corresponding transitions. The numerical simulations of the final model are illustrated in Figure 7, with varying parameter values. In the model, it can be observed that the two important agent types U and V initially increase, approach a peak, then decrease and finally reach 0, hence the issue vanishes. The numerical simulations shown in Figure 7 are also the evidence of the situation. For the dynamics of U i.e., increasing, reaching peak and decreasing has a relationship with S that has been extracted. Also, it has been determined that for what minimum S value the U increases, i.e., what is the condition for U to outbreak?  As the proposed model is complex and comprised of many parameters, for outbreak of U and V, the regression relation has been determined. In Figure 7 (a-e), for various peak values of U, the necessary conditions for U outbreak have been given.
\\
 \begin{figure}
   \centering
       \subfigure[Peak U at $ t=14.5, S_{t}= 0.261964 \approx -0.0293α+0.3268 \frac{1}{\xi}+0.1718 $]{
          \includegraphics[width =50mm]{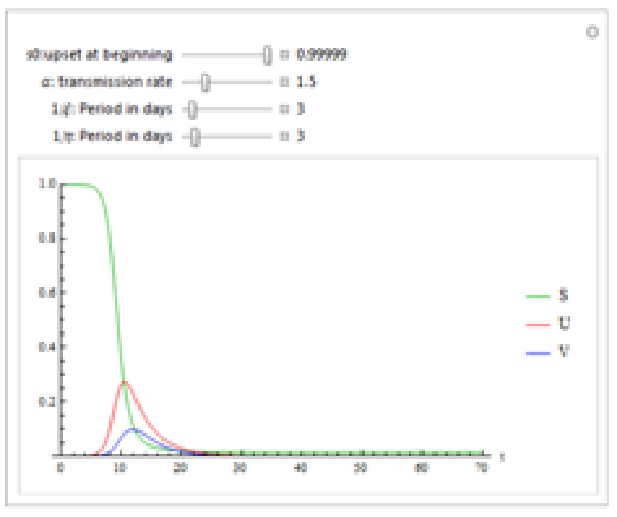}
          }
           \subfigure[Peak U at $ S_{t}= 17.11, S_{t} = 0.3397\approx -0.0293α+0.3268\frac{1}{\xi}+0.1718 $]{           
          \includegraphics[width =50mm]{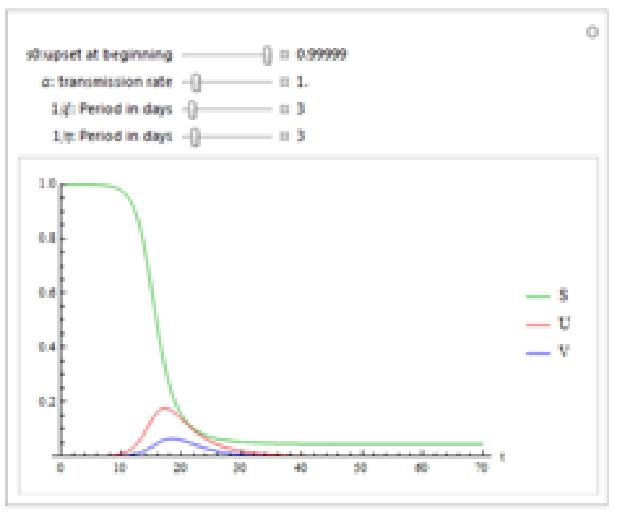}
        }\\ 
        \subfigure[Peak U at $t = 5.4595, S_{t} = 0.0986747 \approx -0.0293α+0.3268\frac{1}{\xi}+0.1718$]{
            \includegraphics[width =50mm]{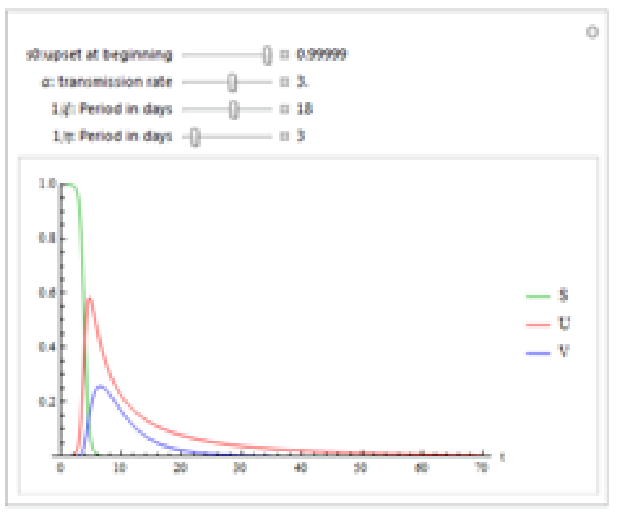}
        }%
          \subfigure[$ (d)Peak U at t=  5.61311, St = 0.13463≈ -0.0293α+0.3268(1/ξ)+0.1718 $]{
            \includegraphics[width =50mm]{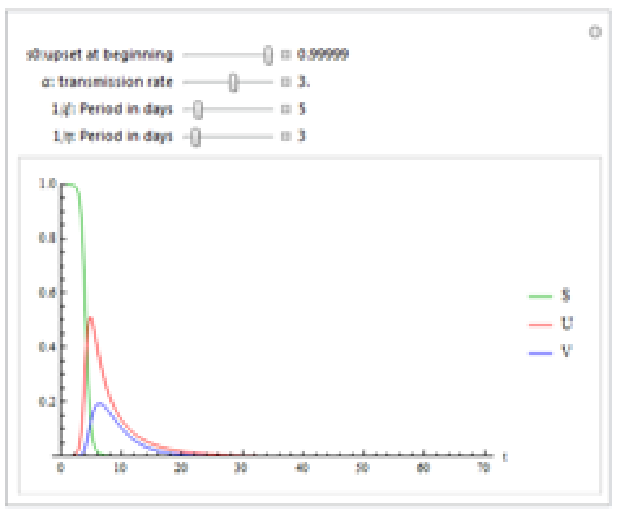}
        }%
      \\
        \subfigure[Peak U at $ t = 8.50755, S_{t} = 0.170486 \approx -0.0293α+0.3268\frac{1}{\xi}+0.1718 $]{
                    \includegraphics[width =50mm]{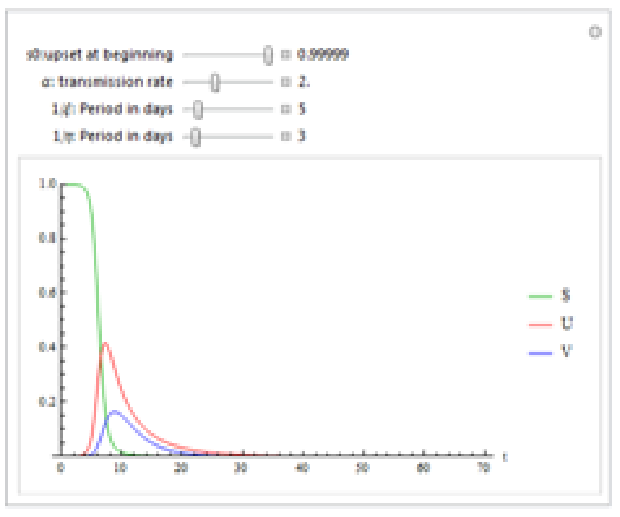}
                }%
    \caption{ (a-e). Numerical simulations of the proposed model with varying peak values of U agents and its relation to $ S_{t} $ }  
\end{figure}
Figure 7 presents the numerical simulations of the proposed five-agent model with the conditions of U outbreak with respect to S; while Figure 8 illustrates the density curves of the agent types S, U, and V  and conditions for V outbreak with respect to U. The overall range of parameter values used in the simulations are given in Table 1.
\begin{figure}
 \centering
 \subfigure[Peak V at $t = 10.3442 $, while $ U_{t}= 0.272683 \approx 0.0283 \alpha -0.4061 \frac{1}{\xi} + 0.429 \frac{1}{\eta} +  0.174 $]{
    \includegraphics[scale =1] {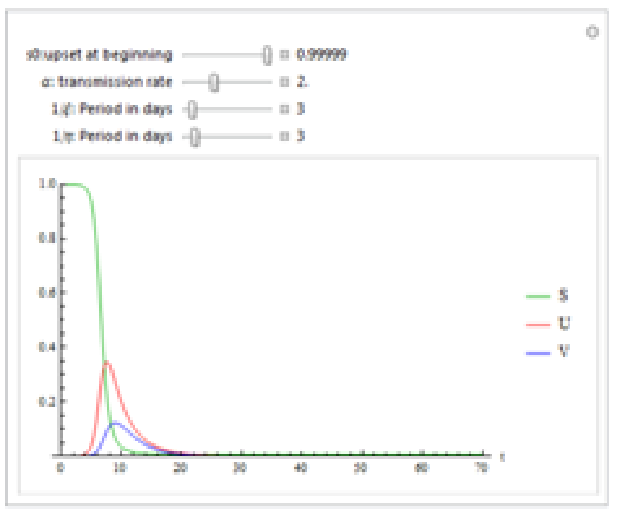}
  }
  
  \subfigure[Peak V at $ t= 9.97177 $, $ U_{t}= 0.392703\approx 0.0283 \alpha -0.4061  \frac{1}{\xi} + 0.429 \frac{1}{\eta} +  0.174 $]{
      \includegraphics[scale =1] {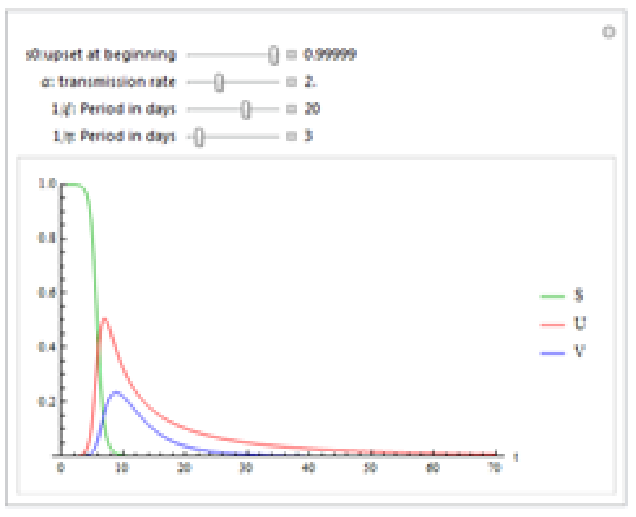}
    }
    
   \subfigure[Peak V at $t= 12.0491 $ and $ U_{t}= 0.274817\approx 0.0283 \alpha -0.4061 \frac{1}{\xi} + 0.429 \frac{1}{\eta} +  0.174 $]{
        \includegraphics[scale =1] {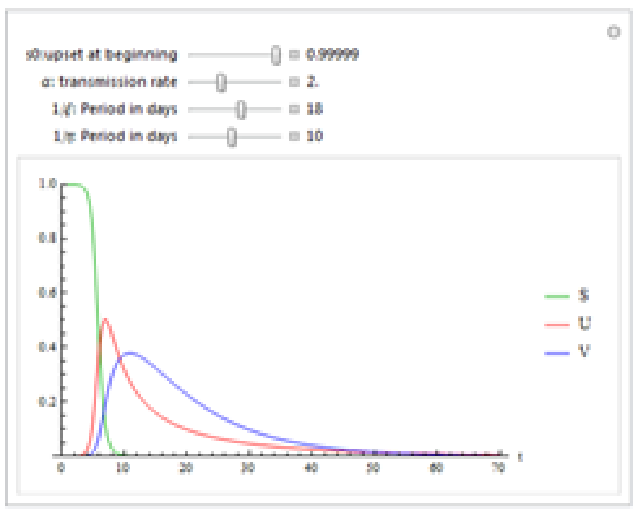}
        }
 \caption{Peak of V with respect to U}
\end{figure} 

Figure 8 presents the dynamics of V agents in the proposed model. In Figure 8 we have used varying values of $ \alpha $ , $ \xi $ and $ \eta $ which affect the peak value and outbreak of the V.
In Figure 7 and Figure 8, we determine the outbreak for U and V respectively, which is estimated using regression relation and is described in following sub-section.
\subsubsection{Outbreak U using regression}
As the value of U is dependent mostly on the value of S, to determine the outbreak of U, we have considered the state of S. A regression relation has been established between U and S, which shows that how U increases, decreases or attains the maximum value in relation to S. In order to determine regression relation, we first extracted peak values of U for various parameter values and established a regression relation between the peak of U and $ S_{t} $. The two parameters that strongly affect the value of U are $ \alpha $ and $ \xi $, so we have taken varying values of these two parameters while rest of the parameter values have been kept constant as given in Table 1.
\\ 
The condition when U is at peak is given by:
\\
\begin{center}
$
S_{t} \approx -0.0293+0.3268\left(\frac{1}{\xi}\right) + 0.1718
$
\end{center}
The necessary condition for U to outbreak is:

\begin{center}
$
S_{0} > -0.0293+0.3268\left(\frac{1}{\xi}\right) + 0.1718
$
\end{center}
The condition for which U begins to decrease, means outbreak does not occur is given by 
\begin{center}
$
S_{0} < -0.0293+0.3268\left(\frac{1}{\xi}\right) + 0.1718
$
\end{center} 
The first condition determines peak U; the second condition holds when outbreak is likely to occur, while third condition holds when outbreak is not likely to occur.
The regression relation established for outbreak has the following properties, which are usually used for estimating the quality of regression relation.
Correlation coefficient = 0.9057, Mean absolute error = 0.0168, Root mean squared error  0.0238.
Correlation coefficient is computed using Eq. 12
\begin{equation}
Correlation Coefficient=\frac{n\sum x y-\left( \sum x\right) \left( \sum y\right) }{\sqrt{n\left( \sum x^{2}\right) -\left( \sum x\right)^{2}}\sqrt{n\left( \sum y^{2}\right) -\left( \sum y\right)^{2}}}
\end{equation}
Where n is the total number of observations, x is observed value and y is computed value. The regression relation is considered to be a good approximation of observed values when coefficient correlation is near +1 and greater than 0.5. The other measures used for estimating quality of regression relation are mean absolute error and root mean squared error. The values near 0 are considered to be good approximations of observed values. 
\subsubsection{Outbreak V using regression}
Like the U agents, the V agents also increase first, reach a peak, and finally decrease and vanish. It is therefore needed to find under what conditions the V increases, when it reaches peak, and when it decreases. The V agents are dependent on the U agents; we have determined regression relation to find out condition for outbreak. The first condition for which V reaches peak is given by: 
\\
\begin{center}
$
U_{t}\approx 0.0283\alpha - 0.4061 \frac{1}{\xi} + 0.429\frac{1}{\eta} + 0.174
$
\end{center}
Thus, the necessary condition for V to outbreak is 
\\
\begin{center}
$
U_{0} > 0.0283\alpha - 0.4061 \frac{1}{\xi} + 0.429\frac{1}{\eta} + 0.174
$
\end{center}
and when this con ``dition holds, V increases; otherwise, it decreases. The accuracy of the regression relation has been determined by three measures that has been used for the quality of U outbreak. These measures include coefficient correlation, mean absolute error, and root mean squared error. 
The regression relation for the outbreak V has following properties:  Coefficient correlation =0.901, Mean absolute error = 0.0309, and Root mean squared error = 0.0396. 
After analyzing the dynamics of the model, we will discuss the final stage of the issue, by determining the total population who ever became aware of the issue.
\section{Final Stage of Ending Dynamics}
When the hatred ``bursting" dynamics cease to end, then only the I, R, and S agents exist. The agents R are those who once were upset or violent but now have been relaxed, while the agents I are those who were aware about the event but did not take any interest in it. The density of the agents I and R at the end determines the total population who ever became aware of the event. The final condition $S_{\infty} = 1 - R_{\infty} - I_{\infty} $ is used to compute the final size of the hatred ``bursting" event aware population. Figure 9 illustrates the combined density of the agents I and R from the beginning until the end of the dynamics.
\\
Figure 9 shows the final state at $t_{\infty}$ when most of the issue-sensitive population was aware of the issue but everyone had become relaxed or was immunized. It may be noted that near t  the sum of the R and I agents approaches 1. In Figure 9, we have used a varying values of $\alpha$, for which the contact between the U and the S agents resulted in new U agents per unit time.
\begin{figure}
\begin{center}
\includegraphics{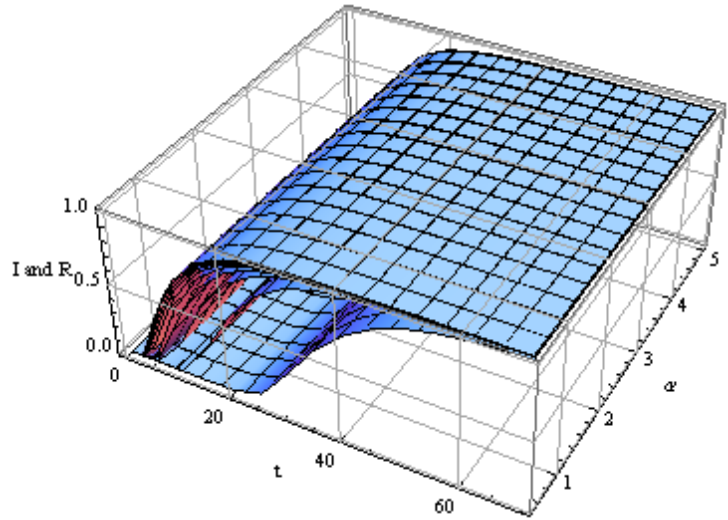}
\caption{Final state of issue awareness on varying rates of $\alpha$}
\end{center}
\end{figure}
\section{Conclusion and Future Work}
In the present research, we explored the growth of population awareness and violence against the hatred ``bursting" events that cause upset and violence among certain groups in a population. We described social behavior of the population using a five-state model by deriving differential equations. We presented numerical simulations of the model  using different initial conditions and various sets of parameters' values.
\\
The outbreak conditions of violence have also been determined. Violent reactions to an issue increase as the upset agents have well motivated the population who then become violent. The mathematical model shows how the upset and violent agents increase at first, then decrease and finally vanish with time. We also have shown why using a basic three-state model of epidemic to hatred contagion is insufficient requiring indeed  a five-state model. 
\\ 
In the future, we plan to further extend the model by predicting the potential victims of violence in the midst of such drastic issues with respect to geographical locations. Once the potential victims are determined, the model would map the potential victims and violent agents on the 2D lattice using percolation theory and determine the likelihood of victimization of the potential victims. We also plan to further investigate the hatred events in order to determine the severity of the issues by keeping in view the issue-sensitive population, the range of population in various geographic locations affected by the issue and some other parameters. 

To conclude, our study may shed a new light to comprehend terror activity and provide some hint on how to limit the spreading of violence within populations globally sensitive to specific world issues. Nonetheless, it should be noted that we are dealing with models which are not the reality although they might well to some extent help to grasp the reality.

\section*{Acknowledgement}
This work was supported in part from a convention DGA-2012 60  0013 00470 75 01 


\end{document}